\documentclass{moriond}

\usepackage{amssymb}
\usepackage{amsmath}
\usepackage[absolute]{textpos}

\def\be{\begin{equation}}
\def\ee{\end{equation}}
\def\bea{\begin{eqnarray}}
\def\eea{\end{eqnarray}}

\newcommand{\Photo}{}

\begin{document}
\vspace*{4cm}
\title{The Spectra of IceCube Neutrino Candidate Sources}

\author{Martina Karl}

\address{Technische Universit{\"a}t M{\"u}nchen, TUM School of Natural Sciences, Physics Department, 
James-Frank-Str. 1, D-85748 Garching bei M{\"u}nchen, Germany \\
European Southern Observatory, Karl-Schwarzschild-Str. 
2, D-85748 Garching bei M\"unchen, Germany}

\maketitle\abstracts{
We present recent work on the Spectra of IceCube Neutrino (SIN) candidate sources project. We defined a selection of candidate neutrino sources by identifying blazars in the vicinity of IceCube's highest-energy neutrinos. We now want to shed light on these source candidates' nature, starting at their redshift, continuing with their black hole masses, their variability, and, for the first time, a combined photon-neutrino spectral energy distribution. We present their hybrid spectral energy distributions (combining photon and neutrino fluxes), investigate the sources' variability in the near-infrared, optical, X-ray, and $\gamma$-ray bands, and compare the variability with a blazar sample of non-candidate sources. Furthermore, we search for flares at the arrival time of the high-energy neutrinos.}

\section{Introduction}

The IceCube Neutrino Observatory detects a flux of high-energetic astrophysical neutrinos. 
Some of these neutrinos are used to alert telescopes of potentially interesting astrophysical phenomena and trigger follow-up observations~\cite{Aartsen_2017}. These neutrino events are called ``alert events''. One alert event and its multi-wavelength follow-up observations identified the first non-stellar neutrino source: the blazar TXS~0506+056~\cite{icfermi}. Blazars are one of the most known powerful non-transient sources and the identification of TXS~0506+056 as a neutrino source motivates the connection of IceCube alert events and blazars.

Ref.~\cite{Giommi:2020hbx} searched for a correlation between $\gamma$-ray detected blazars and IceCube alert events. Out of 47 blazars in the vicinity of IceCube alert events, they found an excess of $16\pm4$ intermediate- and high-synchrotron peaked blazars (IHBLs) with a significance of 3.2$\,\sigma$. 

This triggered the project ``The Spectra of IceCube Neutrino Candidate Sources'' (SIN), where we investigate these candidate neutrino source blazars and search for specific properties that make some of them likely neutrino emitters. 

\section{Optical Spectroscopy}

\begin{figure}
\begin{center}
    \begin{minipage}{0.49\textwidth}
    \includegraphics[width=\textwidth]{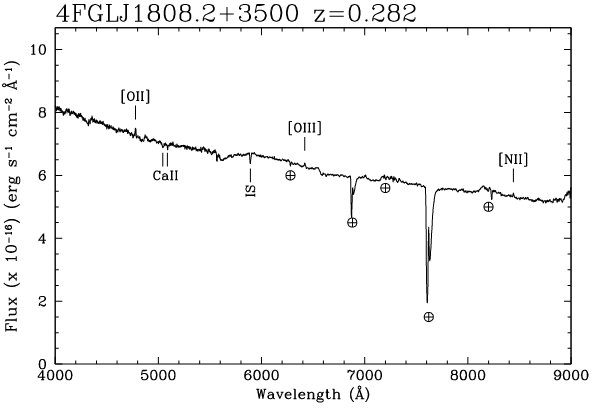}
    \end{minipage}
    \begin{minipage}{0.5\textwidth}
    \includegraphics[width=\textwidth]{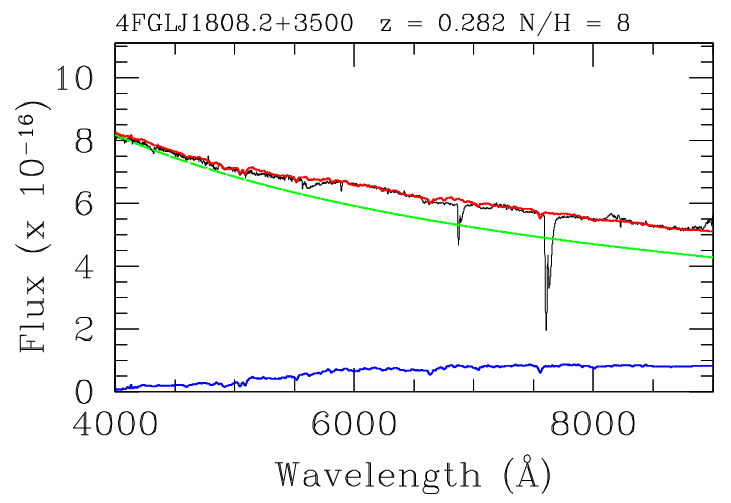}
    \end{minipage}
    \caption{Example spectrum and spectral decomposition of CRATES~J180812+350104. Figures are taken from ref.~\protect\cite{Paiano:2023nsw}. \textbf{Left:} De-reddened and flux-calibrated spectrum obtained by the Large Binocular Telescope. Telluric bands are marked with $\oplus$. \textbf{Right:} Spectral decomposition with the elliptical galaxy template for the host galaxy in blue, the power-law component of the jet in green, and the resulting fit in red. The nucleus-to-host ratio is given at the top.}
    \label{fig:spectral_decomposition}
    \end{center}
\end{figure}

It is crucial to know an object's redshift for a proper source characterization. Out of the 47 neutrino source candidate blazars, 28 had unknown redshifts. An observing campaign was started and as a result, the missing 28 redshifts were determined \cite{Paiano:2021zpc,Paiano:2023nsw}. This was challenging in some cases, since blazars of the class ``BL Lacs'' display few to no spectral features. For some objects, different spectral components had to be fitted (such as the galaxy contribution and the power-law component from the jet) in order to identify spectral lines and determine the redshift (see Fig.~\ref{fig:spectral_decomposition}).  

\section{Source Characterization}
A complete source characterization is now possible with all redshifts known. One major goal is to identify objects that are of a similar nature to TXS~0506+056. Even though TXS~0506+056 was first classified as a BL Lac, closer investigation of the source's spectrum has revealed emission lines common for Flat Spectrum Radio Quasars (FSRQs)~\cite{Padovani2019}. This indicates that the object has a broad line region and its broad emission lines are outshone by its powerful jet~\cite{Padovani2019}. Hence, TXS~0506+056 is not an ordinary BL Lac, but a \textit{masquerading BL Lac}~\cite{10.1093/mnras/stt305}. 

Since TXS~0506+056 is an identified neutrino source, we search for similar blazars, i.e. masquerading BL Lacs. Identifying masquerading BL Lacs depends on various quantities~\cite{Padovani:2021kjr,Paiano:2023nsw}. The main criterion used is a comparison of the radio power at 1.4~GHz, $P_{1.4\mathrm{GHz}}$, to the emission line power $L_\mathrm{O_\mathrm{II}}$, where possible. Another criterion is  $P_{1.4\mathrm{GHz}} > 10^{26}$~W~Hz$^{-1}$, indicating a high excitation of material around the black hole. There are more criteria~\cite{Padovani:2021kjr,Paiano:2023nsw}, some of which rely on relatively uncertain quantities such as the black hole mass or the accretion luminosity. 

\begin{figure}[h!]
\begin{center}
    \begin{minipage}{0.42\textwidth}
    \includegraphics[width=\textwidth]{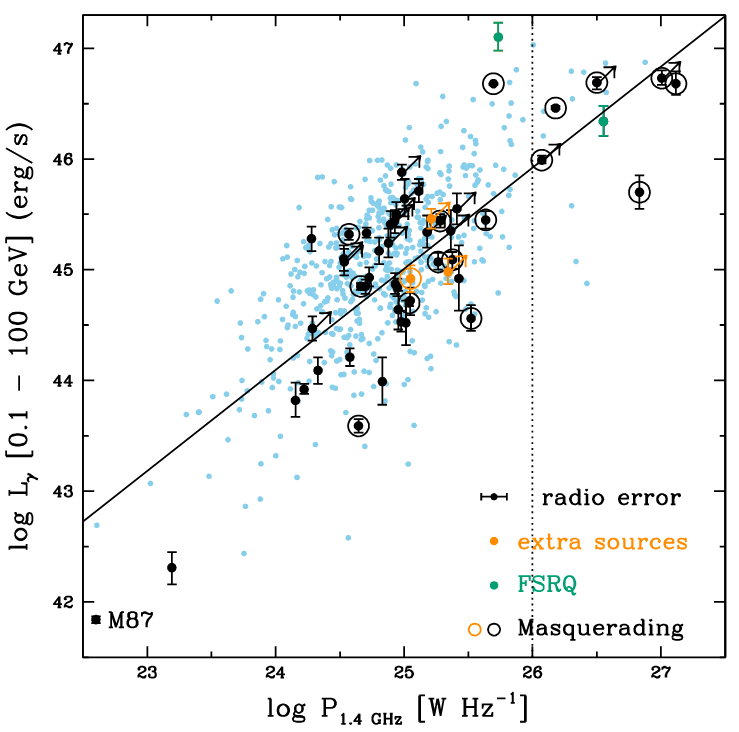}
    \end{minipage}
    \begin{minipage}{0.425\textwidth}
    \includegraphics[width=\textwidth]{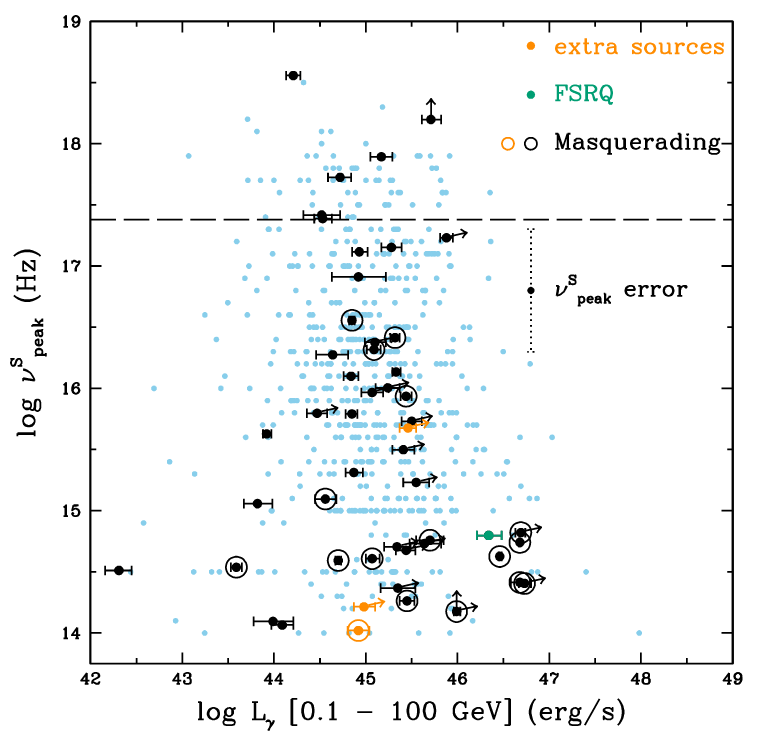}
    \end{minipage}
    \caption{The 47 neutrino source candidate blazars (black dots). Orange points show extra sources associated with alert events detected after the publication of ref.~\protect\cite{Giommi:2020hbx}. Masquerading sources are marked with a circle, FSRQs are shown in green. The small light blue points refer to a blazar control sample. 
    Figures from ref.~\protect\cite{Paiano:2023nsw}. \textbf{Left:} $\gamma$-ray luminosity, $L_{\gamma}$, vs. radio power, $P_{1.4\mathrm{GHz}}$. The solid line shows the linear best fit of $L_{\gamma} \propto P_{1.4\mathrm{GHz}}$. The vertical dotted line shows the threshold $P_{1.4\mathrm{GHz}} = 10^{26}$~W~Hz$^{-1}$. \textbf{Right:} $\nu^\mathrm{S}_\mathrm{peak}$, vs. $L_{\gamma}$. Sources above the dashed line are extreme objects.
    }
    \label{fig:masquerading}
    \end{center}
\end{figure}

An example of source classification is shown in the left panel of Fig.~\ref{fig:masquerading}. Sources with $P_\mathrm{1.4GHz} > 10^{26}$~W~Hz$^{-1}$ are identified as masquerading BL Lacs. The right panel shows the location of the synchrotron peak, $\nu^\mathrm{S}_\mathrm{peak}$, 
vs. $L_{\gamma}$. Extreme sources have $\nu^\mathrm{S}_\mathrm{peak} > 2.4 \times 10^{17}$~Hz. Masquerading sources appear to have lower $\nu^\mathrm{S}_\mathrm{peak}$, agreeing with ref.~\cite{10.1093/mnras/stt305}. Both panels compare our 47 sources with a background blazar population. We find no significant differences between the two groups. 

\begin{figure}[h!]
    \begin{minipage}{0.5\textwidth}
    \includegraphics[width=1\linewidth]{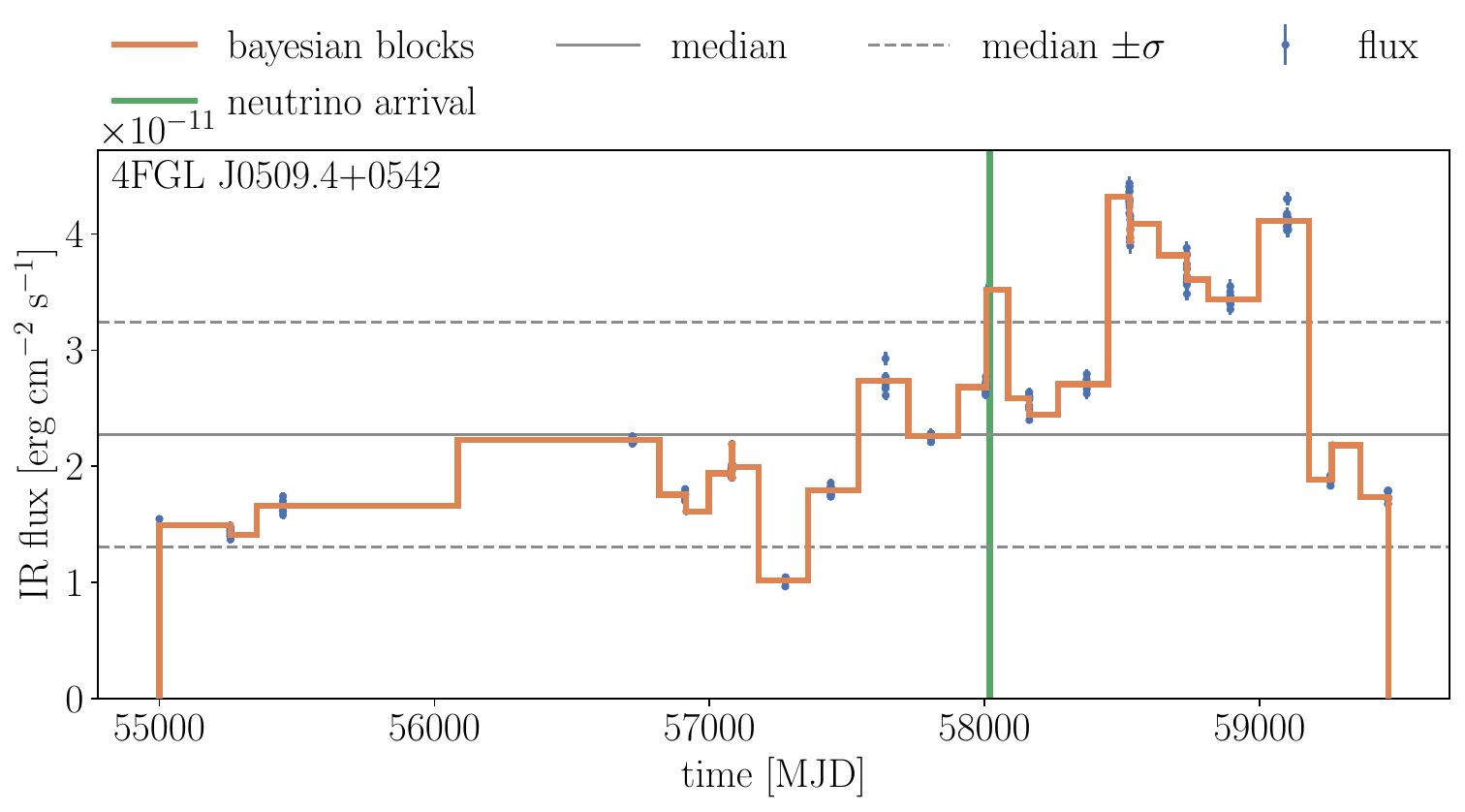}
    \end{minipage}
    \hfill
    \begin{minipage}{0.5\textwidth}
    \includegraphics[width=1\linewidth]{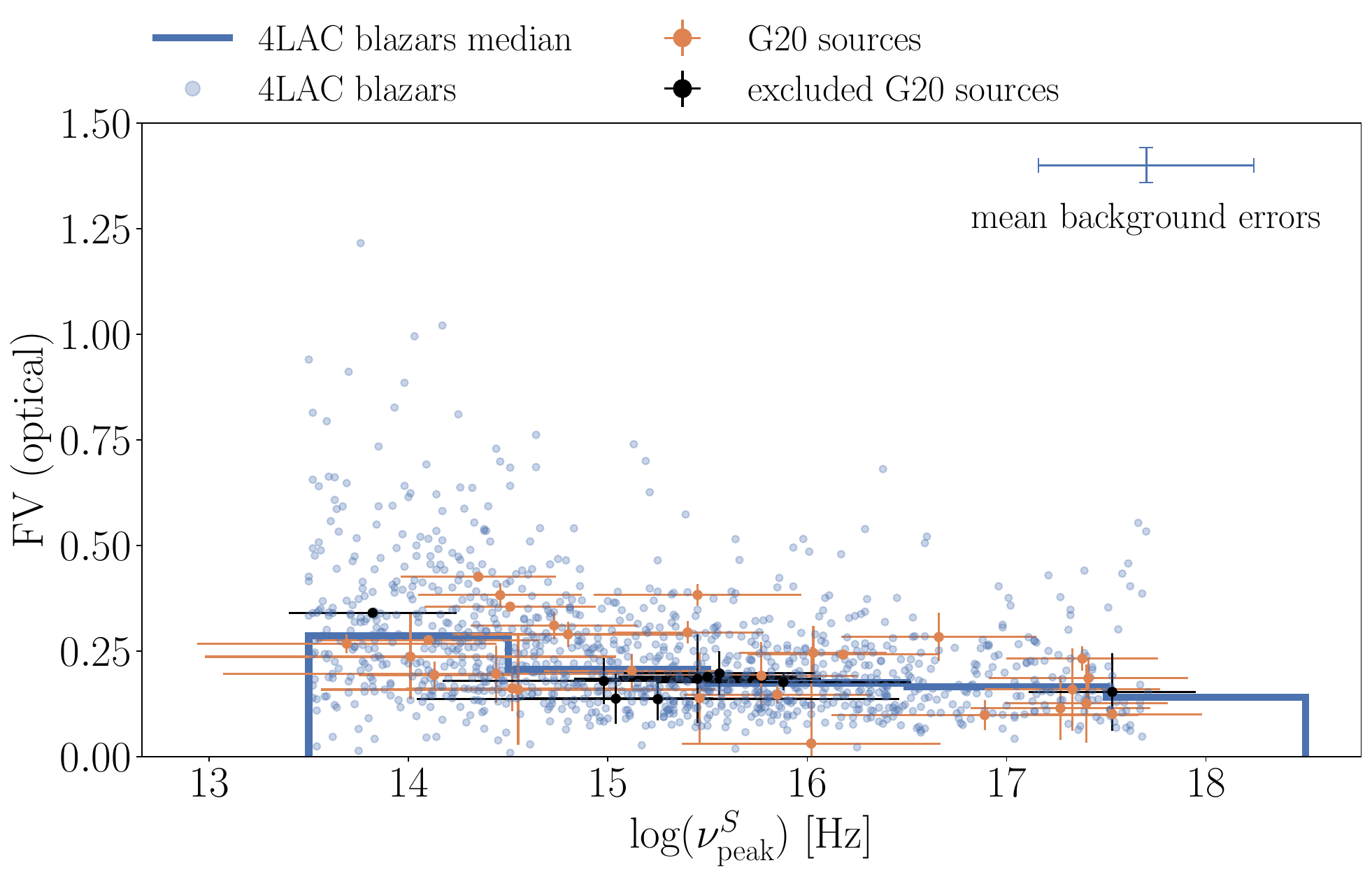}
    \end{minipage}
    \caption{Figures taken from ref.~\protect\cite{Karl:2023huw}. \textbf{Left:} Infrared (IR) lightcurve of TXS~0506+056. Flux bins determined by Bayesian blocks are shown in orange. The neutrino alert arrival time is marked as the vertical green line. The horizontal lines mark the median (solid) $\pm$ the standard deviation (dashed). \textbf{Right:} FV vs. $\nu_\mathrm{peak}^S$. Our sources (orange dots, labeled ``G20 sources'') are plotted on top of a background sample (light blue dots, labeled ``4LAC blazars''). The blue histogram shows the median FVs for the background distribution in bins of $\log \nu_\mathrm{peak}^S$. Sources that do not satisfy updated alert criteria are marked in black.}
    \label{fig:FV_optical}
\end{figure}

In the case of TXS~0506+056, the alert event arrived while the source was flaring~\cite{Aartsen_2017}. We checked if we observe the same pattern with our other sources~\cite{Karl:2023huw}. Hence, we created near-infrared (NEOWISE), optical (ZTF), X-ray (Swift-XRT), and $\gamma$-ray (Fermi-LAT) lightcurves (see, for example, the left panel of Fig.~\ref{fig:FV_optical}). We found no source with flares spanning all bands coincident with alert events apart from TXS~0506+056~\cite{Karl:2023huw}. However, many sources have not been observed at the alert arrival time so the source state at alert arrival remains unknown. This emphasizes the importance of immediate multi-wavelength follow-up of neutrino alerts.

As a next step, we investigated the variability of our source candidates in IR, optical, X-ray, and $\gamma$-rays~\cite{Karl:2023huw}. For this, we calculated the fractional variability (FV), a measure of how much fluxes differ from the average flux variance, following ref.~\cite{Vaughan2003}. 
%
We furthermore updated the selection to only include blazars associated with high-energy tracks that satisfy improved alert criteria published by IceCube~\cite{Abbasi_2023}. This means that from the original 47 sources, 34 remain in our updated selection. The right panel of Fig.~\ref{fig:FV_optical} shows the fractional variability distribution in the optical with data from the Zwicky Transient Facility (ZTF) for our sources and a control sample of blazars in the 4LAC-DR3 catalogue~\cite{Ajello_2020,Lott_2020}. For all investigated wavelengths we find no significant deviation from a background control blazar sample. 

\begin{figure}[h!]
    \begin{minipage}{0.5\textwidth}
    \includegraphics[width=\textwidth]{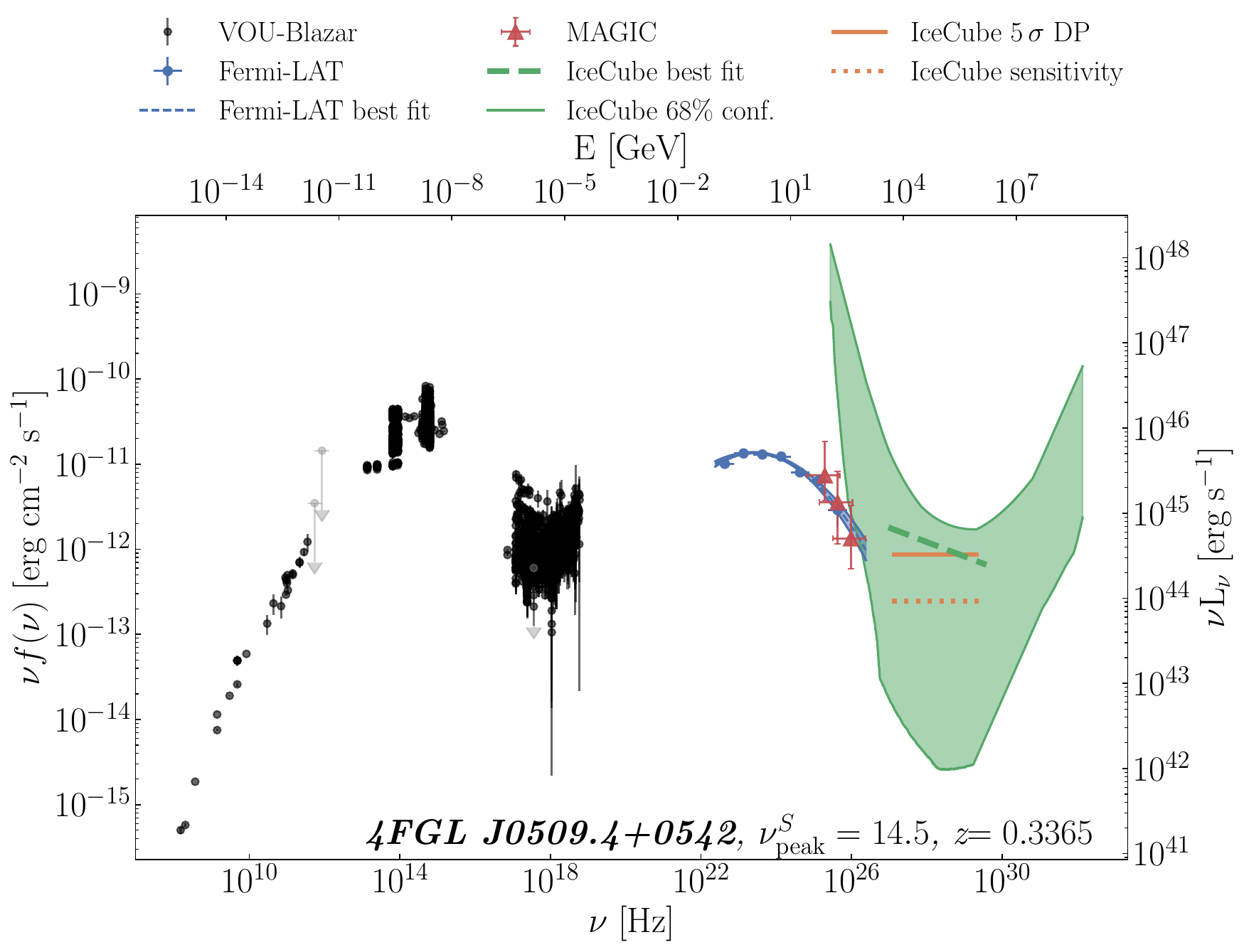}
    \end{minipage}
    \begin{minipage}{0.5\textwidth}
    \includegraphics[width=1\linewidth]{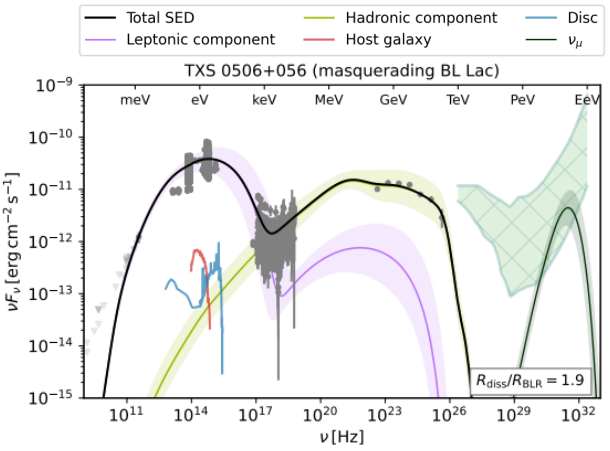}
    \end{minipage}
    \caption{Hybrid multi-wavelength and neutrino SEDs for TXS~0506+056. The green bands show neutrino fluxes agreeing with public IceCube data on a 68\% level. \textbf{Left:} Multi-wavelength SED (black, blue, and red points) and neutrino emission assuming a power-law energy spectrum (dashed green line). Figure from ref.~\protect\cite{Karl:2023huw}. \textbf{Right:} Modeled multi-wavelength and neutrino emission of TXS~0506+056 (black lines). The jet comprises leptonic (purple) and hadronic (light green) components. The galaxy contributions are displayed in red, and the accretion disc, broad line region, and dust torus are shown in blue. The modeled neutrino flux is compatible with IceCube data if the spectrum's peak lies within the green cross-hatched band. Figure from ref.~\protect\cite{SINV}.}

    \label{fig:sed}
\end{figure}

We determined the sources' neutrino emission and show hybrid multi-wavelength and neutrino spectral energy distributions (SEDs)~\cite{Karl:2023huw} (left panel of Fig.~\ref{fig:sed}). We set 68\% confidence level lower limits for 11 out of 34 sources and upper limits for the remaining ones~\cite{Karl:2023huw}. When looking for $\gamma$-ray fluxes being of the same order of magnitude as the neutrino fluxes, we found a correlation on a $2.2\,\sigma$ level~\cite{Karl:2023huw}.

As a next step, we updated the neutrino emission model to follow state-of-the-art lepto-hadronic models and fit the models to the data~\cite{SINV} (see right panel of Fig.~\ref{fig:sed} for an example and the updated peaked neutrino spectrum). The modeled emission of masquerading BL Lacs shows a strong neutrino component~\cite{SINV}.  

\section{Summary and Outlook}

The SIN project investigates a selection of blazars associated with high-energy neutrinos where approximately a third are expected neutrino sources. We aim to characterize the sources and identify features that distinguish some blazars as neutrino emitters.
So far, we have not found a significant difference between the objects studied within the SIN project and blazars satisfying the selection criteria but not being associated with high-energy neutrino events. 

\section*{Acknowledgments}
The results discussed here were obtained as part of the SIN collaboration \cite{Paiano:2021zpc,Padovani:2021kjr,Paiano:2023nsw,Karl:2023huw,SINV}. This work is supported by the Deutsche Forschungsgemeinschaft (DFG, German Research Foundation) through grant SFB 1258 ``Neutrinos and Dark Matter in Astro- and Particle Physics''.

\end{document}